%% file: CorFluc06_Lacey3.tex
\documentclass{PoS}
\usepackage{epsfig}
\title{What do elliptic flow measurements tell us about the matter created in the little Bang at RHIC?}

\ShortTitle{What do elliptic flow measurements tell us about the matter created at RHIC?}

\author{\speaker{Roy A. Lacey} and Arkadij Taranenko\\
        Dept. of Chemistry, Stony Brook University, Stony Brook, NY, 11794-3400, USA.\\
        E-mail: \email{Roy.Lacey@Stonybrook.edu}}

\abstract{ Elliptic flow measurements are presented and discussed with emphasis on 
the hydrodynamic character of the hot and dense QCD matter created in 
heavy ion collisions at RHIC. Predictions from perfect fluid hydrodynamics 
for the scaling of the elliptic flow coefficient $v_2$ with eccentricity, 
system size and transverse energy are validated. 
A universal scaling for the flow of both mesons and baryons is observed for a 
broad transverse kinetic energy range when quark number scaling is employed. 
This suggests a new state of nuclear matter at extremely high density and 
temperature whose primary constituents have the quantum numbers of quarks 
and anti-quarks in chemical equilibrium.
The scaled flow is used to constrain estimates for several transport 
coefficients including the sound speed $c_s$, shear viscosity to entropy ratio 
$\eta/s$, diffusion coefficient ($D_c$) and sound attenuation length ($\Gamma$).
The estimated value $\eta/s \sim 0.1$, is close to the absolute lower bound ($1/4\pi$), 
and may signal thermodynamic trajectories for the decaying matter 
which lie close to the QCD critical end point.
}

\FullConference{The 2nd edition of the International Workshop --- Correlations and Fluctuations in Relativistic Nuclear Collisions --- \\
		 July 7-9 2006\\
		 Galileo Galilei Institute, Florence, Italy}
\begin{document}

\section{Introduction}
	Recent experiments at Brookhaven's Relativistic Heavy Ion Collider (RHIC), 
give strong evidence for the creation of locally thermalized hot and dense QCD matter 
in a ``little bang" initiated in ultra relativistic nucleus nucleus 
collisions \cite{Adcox:2004mh}. The energy density $\varepsilon$, achieved a 
short time after the little bang ($\sim 1fm/c$) has been estimated to 
be  $\ge 5.4$~GeV/fm$^3$ \cite{Adcox:2004mh} 
 -- a value significantly larger than the $\sim 1$~GeV/fm$^3$ required for the 
transition from hadronic matter to the high-temperature plasma 
phase of QCD \cite{Karsch:2001vs,Fodor:2001pe}. A confluence of experimental 
results \cite{Adcox:2004mh,Adams:2005dq,Back:2004je,Arsene:2004fa}  
now suggest that this new state of matter bears strong kinship to the hot and dense 
plasma of quarks and gluons (QGP) produced a few micro seconds after the ``big bang" 
that gave birth to our universe some $12-14$ billion years ago. Now, the scientific 
challenge is to explore robust experimental constraints which can establish 
the detailed properties of this QGP. 

	In this contribution, we show experimental validation for; 
(i) the scaling predictions of perfect fluid hydrodynamics for the elliptic  
flow coefficient $v_2$, (ii) the development of elliptic flow in the pre-hadronization 
phase, (iii) the scaling of D meson flow compatible with full thermalization of the 
charm quark and (iv) universal scaling of the flow of both mesons 
and baryons (over a broad transverse kinetic energy range) via quark number 
scaling. Subsequently, we use the scaled flow values to constrain a number of 
transport coefficients.

\section{Thermalization and harmonic flow correlations}
	Local thermalization plays a central role in considerations involving perfect fluid 
hydrodynamics. Consequently, it is important to recap a few of the key 
observables which are consistent with the production of high energy density 
thermalized matter at RHIC \cite{Lacey:2005qq}. In brief;
\begin{itemize}
\item the multiplicity distribution for the same number of participants is independent 
of colliding system size as would be expected from a system which ``forgets"  how it is 
formed. A demonstration of this independence hypothesis has been given 
for Au+Au and Cu+Cu collisions by the PHOBOS collaboration \cite{Roland:2005ei}.
\item A comparison of the measured anti-particle to particle yield ratios ${\bar p}/p$, to  
the predictions of a statistical model
\begin{equation}
\frac{{\bar p}}
{p} = \frac{{e^{ - (E + \mu )/T} }}
{{e^{ - (E - \mu )/T} }} = e^{ - 2\mu /T}, 
\end{equation}
show excellent agreement for a single temperature $T$ and baryon chemical 
potential $\mu$ \cite{Braun-Munzinger:2001ip}.

\item Surprisingly large harmonic flow correlations compatible with the development of 
sizable  pressure gradients in a locally thermalized fluid are observed for a broad selection 
of hadrons comprised of light-  and heavy quark 
combinations ($\pi, K, \mathrm{p}, \Lambda, \Omega, \phi, \Xi, \mathrm{d}, \mathrm{D}, \ldots$)
\cite{Adler:2003kt,Adams:2003am,Adams:2004bi,Adams:2005zg,Sakai:2005qn}. 
We emphasize here that rather strong interactions are required in this fluid to 
generate significant flow for hadrons comprised of a heavy quark, ie. the heavy quark 
relaxation time is much longer than that for the light quarks.

\end{itemize}
	The magnitude of harmonic flow correlations is commonly characterized by the Fourier 
coefficients $v_n$ as \cite{Voloshin:1994mz}:
\begin{equation}
v_n\equiv\langle e^{in(\varphi-\Phi_R)}\rangle
=\langle \cos n(\varphi-\Phi_R)\rangle,
\label{rxn_plane}
\end{equation}
where $\varphi$ is the azimuthal angle of an emitted particle
(measured in the laboratory coordinate system) and $\Phi_R$ is 
an estimate of the azimuth of the reaction plane. A correction factor 
for $v_n$ is usually required to take account of the dispersion of the 
estimated reaction plane \cite{Danielewicz:1985hn,Poskanzer:1998yz,Ollitrault:1997di}. 
The second harmonic coefficient $v_2$ is termed elliptic flow.

	At RHIC energies, it is widely believed that elliptic flow results from hydrodynamic 
pressure gradients developed in an initial ``almond-shaped" collision zone. That is, 
the initial transverse coordinate-space anisotropy of the collision zone is converted, 
via particle interactions, into an azimuthal momentum-space anisotropy. Elliptic flow 
self-quenches due to expansion of the collision zone. Consequently, rapid local 
thermalization is required to achieve relatively large $v_2$ signals. The current 
consensus is that the sizable values observed at RHIC are compatible with the 
hydrodynamic expansion of a near perfect fluid.
%}  

%
\section{Universal scaling and perfect fluid hydrodynamics \label{uv_scaling}}

	Perfect fluid hydrodynamics stipulates that the constituents of a fluid 
flow with the same velocity field. In many hydrodynamic models 
\cite{Ollitrault:1992bk,Heiselberg:1998es,Huovinen:2001cy,
teaney2001, kolb2001, hirano_qm05, Bhalerao:2005mm}, elliptic flow 
results from pressure gradients due to the initial spatial asymmetry 
or eccentricity $\epsilon= (\langle y^2- x^2\rangle)/(\langle y^2+ x^2\rangle)$, 
of the high energy density matter in the collision zone. 
The initial entropy density $S(x,y)$, is often used to perform an average over the 
$x$ and $y$ coordinates of the matter in the plane perpendicular to the 
collision axis. 
For a system of transverse size $\bar R$ ($1/\bar R=\sqrt{1/\langle x^2\rangle+1/\langle y^2\rangle}$), 
this flow develops over a time scale $\sim \bar R/\left\langle c_s\right\rangle$
for matter with sound speed $c_s$ \cite{Bhalerao:2005mm}. Thus, the initial energy density controls 
how much flow develops globally, while the detailed development of the flow 
correlation patterns are governed by $\epsilon$ and $c_s$. Centrality dependent 
differential measurements give clear access to $\epsilon$; access to $c_s$ is less 
direct but is reflected in the magnitude of the flow. 
Lattice QCD calculations indicate that $c_s^2 \approx 1/3$ for $T > 2T_{c}$, but 
drops steeply by more than a factor of six near the ``softest point" 
where $T \approx T_{c}$ ie. the critical temperature \cite{Karsch:2006sm}. Thereafter, it rises 
again in the hadron resonance gas phase to the value $c_s^2 \approx 0.15$ 
for SPS energies. 

\begin{figure}[!h]
%\vspace{.20in}
\begin{tabular}{ccc}\hspace{-.4in}
\begin{minipage}{.6\linewidth}
\includegraphics[width=1.\linewidth]{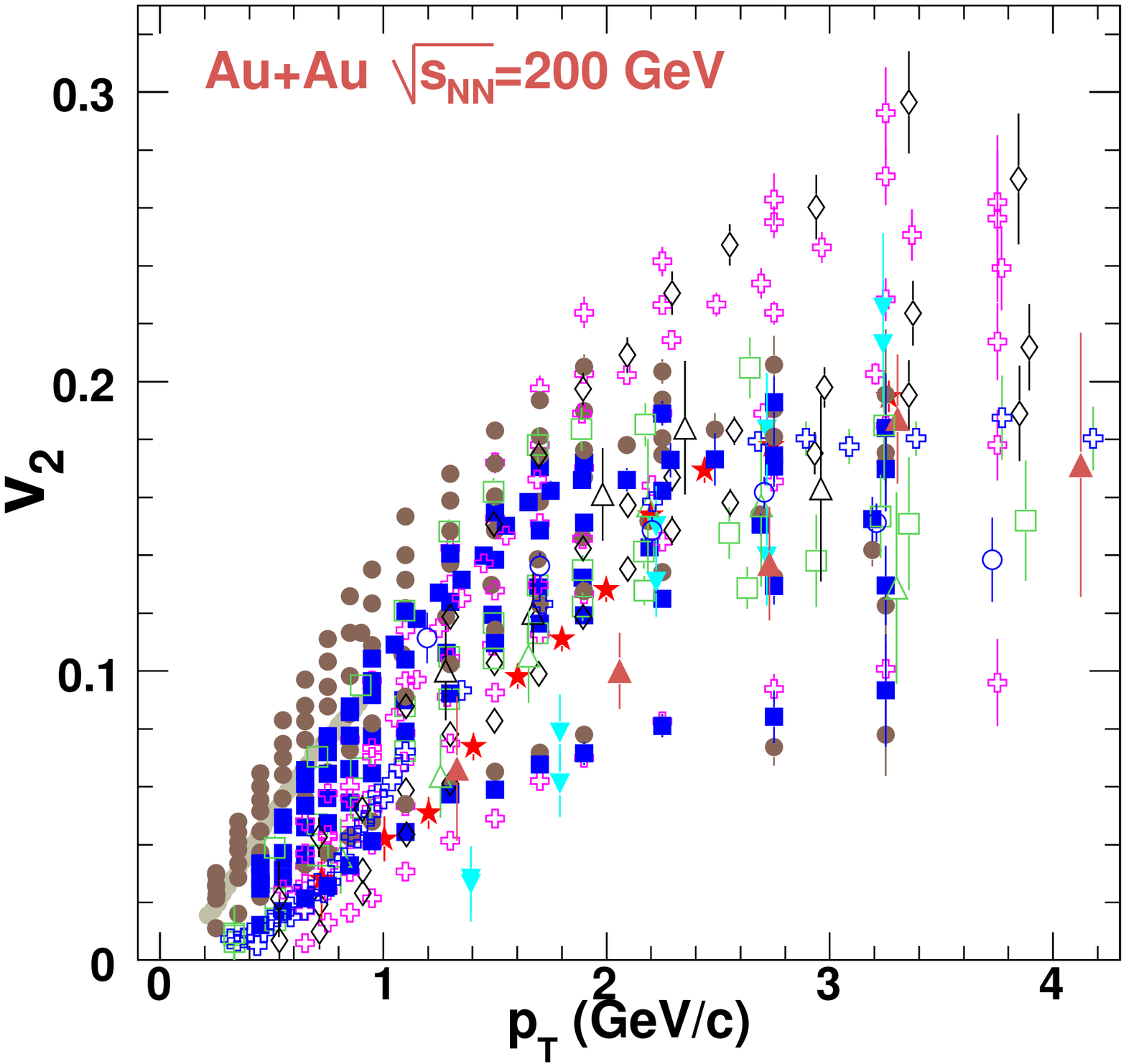}
\end{minipage} &\hspace{-0.2in}
\begin{minipage}{.45\linewidth}
\includegraphics[width=1.\linewidth]{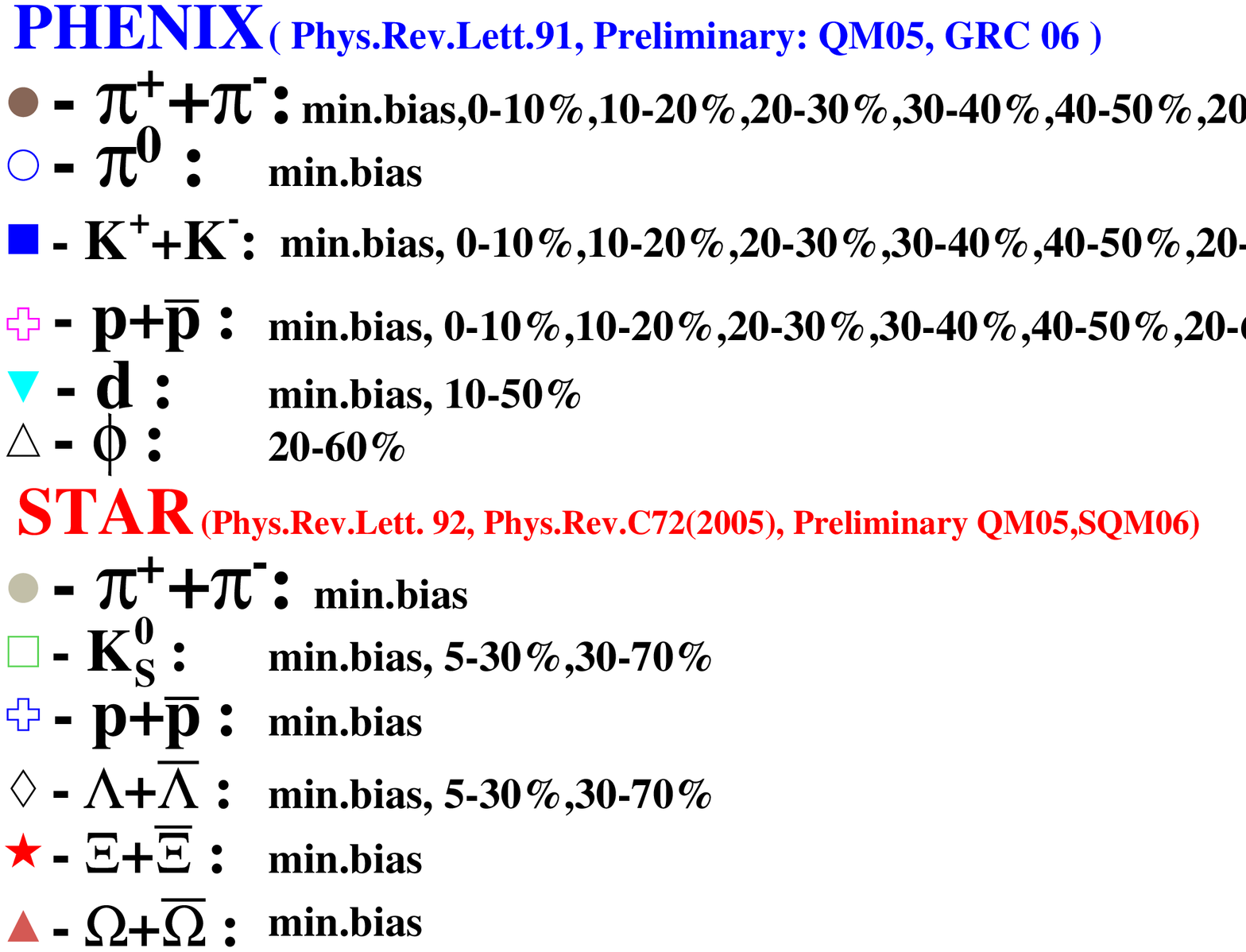}
\end{minipage} &
\end{tabular}
%\vspace{.10in}
\caption{$v_2$ vs. $p_T$ for several particle species produced at midrapidity in central 
and semi-central Au+Au collisions at $\sqrt{s_{NN}} = 200$~GeV.
The centrality and particle species selections are indicated. }
\label{scaling_fig_pt}
\end{figure}

\begin{figure}[!h]
%\vspace{.20in}
\begin{tabular}{ccc}\hspace{-.4in}
\begin{minipage}{.6\linewidth}
\includegraphics[width=1.\linewidth]{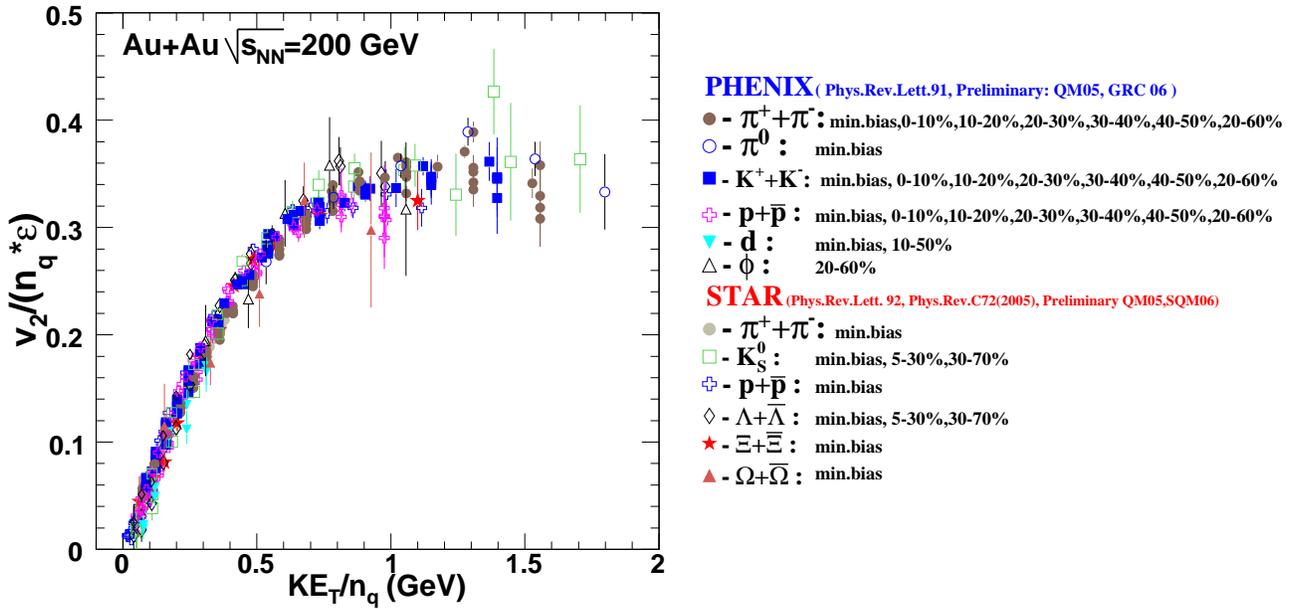}
\end{minipage} &\hspace{-0.2in}
\begin{minipage}{.45\linewidth}
\includegraphics[width=1.\linewidth]{legend.eps}
\end{minipage} &
\end{tabular}
%\vspace{.10in}
\caption{$v_2/n_q \epsilon$ vs. $KE_T/n_q$ for several  
particle species produced at midrapidity in central and semi-central Au+Au collisions 
at $\sqrt{s_{NN}} = 200$ GeV. The data selections are indicated. }
\label{scalling_fig_ketn}
\end{figure}

	A remarkable empirical fact recently reported, is the observation that universal 
scaling for the flow of both mesons and baryons is achieved when $v_2/n_q\epsilon$ 
is plotted vs. $KE_T/n_q$ \cite{Issah:2006qn,Adare:2006ti}. 
Here, $n_q$ is the number of valence quarks ($n_q =$~2, 3 for mesons and baryons 
respectively) and $KE_T=m_t-m$ is the transverse kinetic energy\footnote{The pressure in 
an ideal gas can be seen to be a measure of its kinetic energy density.}. 
Further demonstration of this scaling is illustrated via Figs. \ref{scaling_fig_pt} 
and \ref{scalling_fig_ketn}.  Figure \ref{scaling_fig_pt} shows differential flow 
measurements $v_2(p_T)$, for several particle species produced at mid-rapidity in 
central and semi-central Au+Au collisions at $\sqrt{s_{NN}} = 200$ GeV; they 
span essentially the full range of measurements (several hundred data points) at RHIC. 
The $v_2$ values shown for charged  and neutral pions ($\pi^{\pm}$, $\pi^{0}$), 
charged kaons ($K^{\pm}$), (anti-) protons ($(\overline{p})p$), (anti-) deuterons 
($(\overline{d})d$ and the $\phi$ meson represent recent measurements 
(published and preliminary results) obtained by the PHENIX collaboration 
\cite{Lacey:2005qq,Adler:2005rg,Issah:2006qn,Adare:2006ti}. 
The values for neutral kaons ($K^0_s$), lambdas ($\Lambda+\bar{\Lambda}$), 
cascades ($\Xi+\bar{\Xi}$) and omegas ($\Omega+\bar{\Omega}$) show results from the STAR
collaboration \cite{Adams:2003am,Adams:2005zg,Lamont:2006rc,Oldenburg:2006br}.

The most striking feature of this plot is the large 
magnitude of $v_2(p_T)$ observed for all particle species. 
Fig. \ref{scalling_fig_ketn} shows the scaled results obtained from the 
same data. It indicates that the relatively complicated ``fine structure" 
of $v_2$ (ie. its detailed dependence on centrality, transverse momentum, 
particle type, etc) for particles produced at mid-rapidity  can be scaled 
to a single function.
%
%===============================================================
%

\begin{figure}[tb]
\vspace{.10in}
%\begin{minipage}[t]{0.5\linewidth}
\includegraphics[width=1.\linewidth]{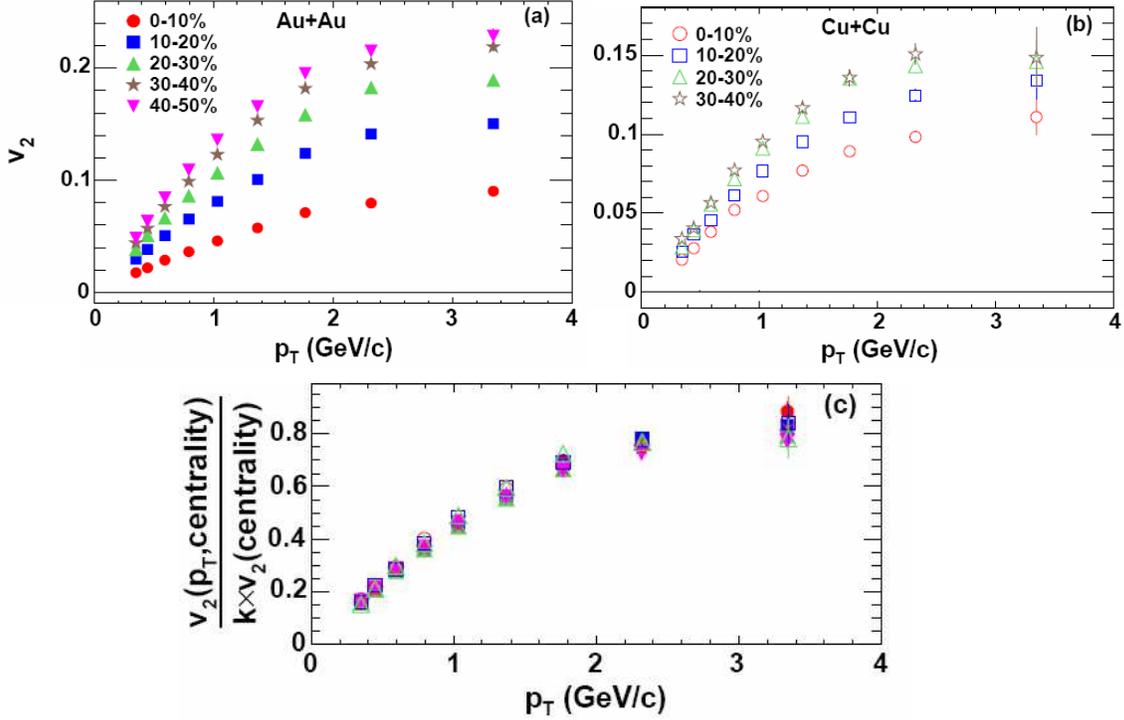}
%\vskip -1.2cm
\caption{ $v_2$ vs. $p_T$ for charged hadrons obtained in Au+Au (a) and
Cu+Cu (b) collisions for several centralities as indicated. Panel (c) 
shows $v_2(centrality, p_T)$ divided by $k$ times the $p_T$-integrated 
value $v_2(centrality)$, for Au+Au and Cu+Cu  ($k = 3.1$ see text 
and Refs. \cite{Issah:2006qn,Adare:2006ti}).  
} 
\label{ecc_scaling}
\end{figure}
\begin{figure}[tb]
%\begin{minipage}[t]{0.5\linewidth}
\includegraphics[width=1.\linewidth]{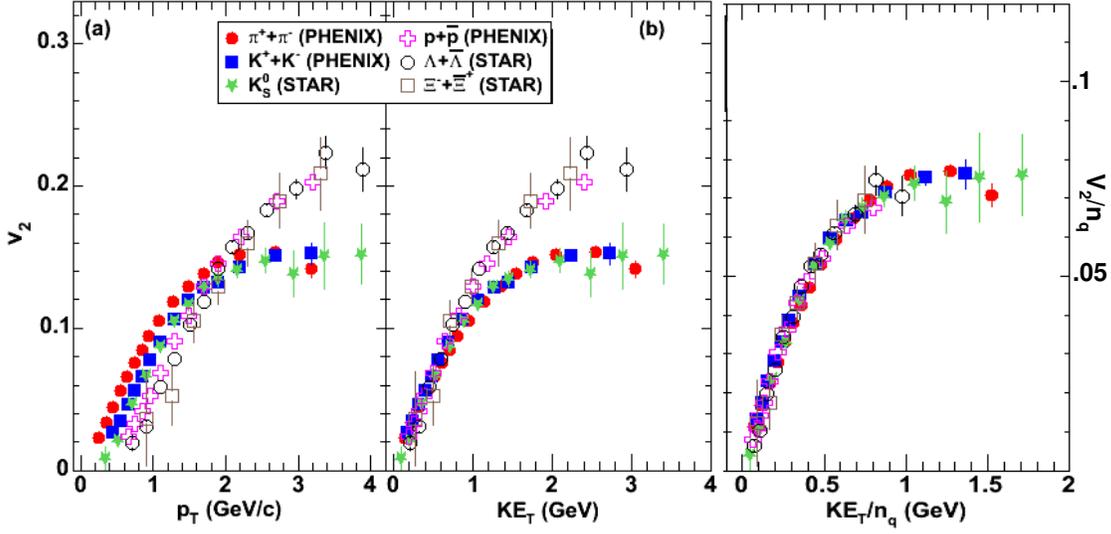}
%\vskip -1.2cm
\caption{ $v_2$ vs. $p_T$ (left panel) and $KE_T$ (middle panel). The 
scaled results in the right panel is obtained by $n_q$ scaling the data 
shown in the middle panel. Results are shown for several particle species 
produced in minimum bias Au+Au collisions at $\sqrt{s_{NN}} = 200$ GeV \cite{Adare:2006ti}. 
} 
\label{scalling_fig_pt-to-ket}
\end{figure}
	On the one hand, this universal scaling validates the predictions of perfect 
fluid hydrodynamics \cite{Heiselberg:1998es,Bhalerao:2005mm,Csanad:2005gv,Borghini:2005kd,Csanad:2006sp}. 
That is; 
\begin{itemize}
\item $v_2(p_T)/\epsilon$ should be independent of centrality; 
\item $v_2(p_T)$ should be independent of colliding system size for a given 
eccentricity;
\item for different particle species, $v_2(KE_T)$ at mid-rapidity should scale 
with the transverse kinetic energy $KE_T = m_T-m$, where $m_T$ is the transverse 
mass of the particle; 
\item $v_4(p_T) \propto v_2^2(p_T)$.
\end{itemize}
On the other hand, it corroborates the $n_q$ scaling expected from quark coalescence 
models \cite{Voloshin:2002wa,Fries:2003kq,Molnar:2003ff,Greco:2003mm} suggesting that quark-related
degrees of freedom are present when elliptic flow develops. 

	These separate aspects of the observed universal scaling are illustrated 
in Figs.\ref{ecc_scaling} and \ref{scalling_fig_pt-to-ket}.
Figures \ref{ecc_scaling}a and \ref{ecc_scaling}b show the differential $v_2(p_T)$ 
for charged hadrons obtained in Au+Au and Cu+Cu collisions respectively. 
The expected increase in $v_2(p_T)$ as collisions become more peripheral and the $p_T$ 
increases \cite{Adcox:2004mh,Adams:2005dq,Back:2004je} is unmistakable. 
Fig. \ref{ecc_scaling}c show the resulting $\epsilon$-scaled values for both 
systems. The eccentricity is obtained from the $p_T$-integrated $v_2$ 
values i.e. $\epsilon = k \times v_2$. A Glauber model estimate of 
$\epsilon$ in Au+Au collisions \cite{Adcox:2002ms} gives $k = 3.1 \pm 0.2$ 
for the cuts employed in the analysis \cite{Issah:2006qn,Adare:2006ti}. 

	These scaled values shown in Fig. \ref{ecc_scaling}c validate the 
expected independence on colliding system size \cite{Lacey:2005qq,Bhalerao:2005mm} 
and suggests that rapid local thermalization \cite{Shuryak:2004cy,Heinz:2001xi} is 
achieved. They also allow an estimate of $c_s$ because the magnitude of $v_2/\epsilon$ 
depends on the sound speed $c_s$ \cite{Bhalerao:2005mm}.  
A recent estimate \cite{Issah:2006qn,Adare:2006ti} gives $c_s \sim 0.35 \pm 0.05$; a value 
which suggests an effective equation of state (EOS) which is softer than that for the high 
temperature QGP \cite{Karsch:2006sm}. It however, does not reflect a strong first order phase 
transition in which $c_s = 0$ during an extended hadronization period. Such an 
EOS could very well be the reason why $v_2(p_T)$ is observed to saturate in Au+Au 
collisions for the collision energy range 
$\sqrt{s_{NN}} = 60 - 200$ GeV \cite{Adler:2004cj}.

	The other aspects\footnote{Validation of the hydrodynamic scaling 
prediction that $v_4 \propto v^2_2$ is presented in Ref. \cite{Lacey:2005qq}. 
It is a non-trivial prediction; the measured differential $v_2(p_T)$ show 
a linear $p_T$ dependence for low $p_T$ pions but $v_4(p_T)$ is quadratic in $p_T$.}
of the observed universal scaling are illustrated 
in Fig.\ref{scalling_fig_pt-to-ket}. The left panel show measurements 
of the $p_T$ dependence of $v_2$  for several particle species. The middle panel 
bears out the expected particle mass scaling when $v_2$ is plotted vs. $KE_T$. 
For $KE_T \geq 1$~GeV, clear splitting into a meson branch (lower $v_2$) and a 
baryon branch (higher $v_2$) occurs. However, both of these branches 
show rather good scaling separately. 
The right panel shows the resulting values obtained after scaling both 
$v_2$ and $KE_T$ (ie the data in the middle panel) by $n_q$.  
This latter scaling is an indication of the inherent quark-like degrees 
of freedom in the flowing matter. That is, we assert that the bulk of the 
elliptic flow develops in the pre-hadronization phase.

\subsubsection{Validation of the pre-hadronic development of elliptic flow}
	Despite robust quark number scaling, one can still ask whether or not the bulk of the 
elliptic flow develops pre- or post-hadronization. This question can be readily addressed 
via careful study of the $v_2$ for the $\phi$ meson. 
The $\phi$ meson is comprised of a strange (s) and an anti-strange ($\bar{s}$) quark
and its interaction with nuclear matter is suppressed according to the 
Okubo-Zweig-Izuka (OZI) rules \cite{Okubo:1963fa}. Therefore, the $\phi$ meson is expected 
to  have a rather small hadronic cross section with non-strange hadrons ($\sim 9$~mb) 
\cite{Shor:1984ui,Ko:1993id,Haglin:1994xu} and this leads to the relatively large mean free 
path $\lambda_\phi$, when compared to the transverse size of the emitting 
system. 

Thus, if elliptic flow was established in a phase involving hadrons interacting 
with their standard hadronic cross sections (post-hadronization), one 
would expect $v_2$ for the $\phi$ to be significantly smaller than that for  
other hadrons (eg. $p$ and $\pi$). An added bonus is that the relatively long 
lifetime ($\sim 45$~fm/c) for the $\phi$ makes its decay inside the 
pre-hadronization medium unlikely and this makes for a cleaner measurement. 
For these reasons, flow measurements for the $\phi$ 
play a crucial role for the  distinction between pre- and post-hadronic 
development of elliptic flow \cite{Xu:2005jt}.
\begin{figure}[tb]
\includegraphics[width=1.\linewidth]{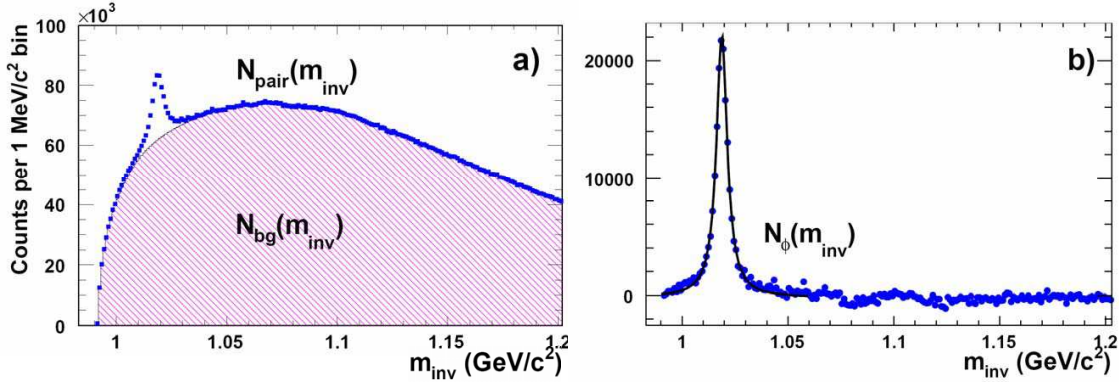}
%\vskip -1.2cm
\caption{ (Color on line)  An example of invariant mass distributions for K$^{+}$K$^{-}$ pairs
identified in the PHENIX detector for $p_{T}^{pair}$=1.6-2.7 GeV/c and reaction centrality 20-60\%:
The left panel shows the distribution for signal plus combinatoric background. The right panel shows 
the distribution after background subtraction.
}  
\label{Phi_invmass}
\end{figure}
\begin{figure}[tb]
\includegraphics[width=1.\linewidth]{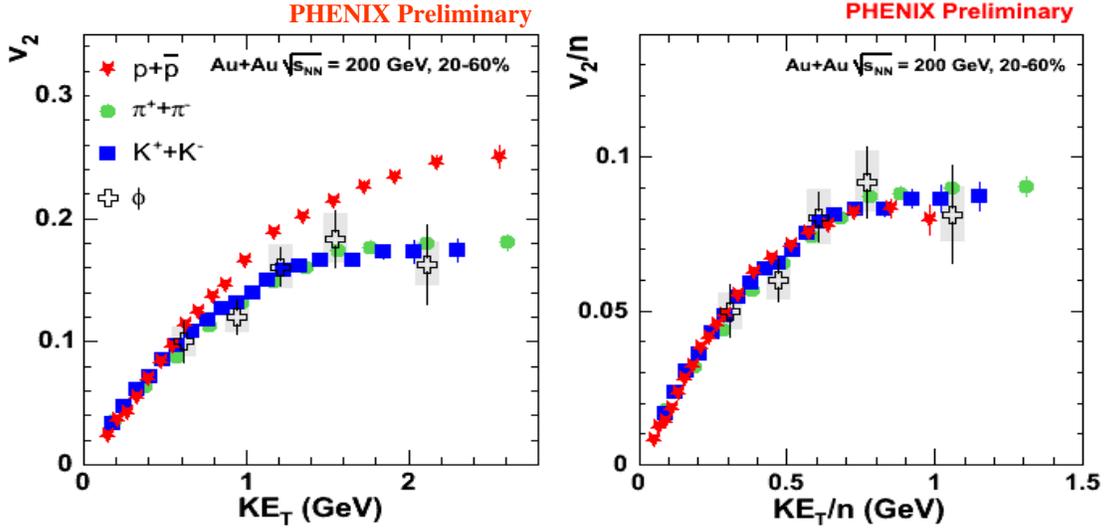}
%\vskip -1.2cm
\caption{ (Color on line) Comparison of $v_2$ vs. $KE_T$ for $\pi, K, p$ and $\phi$ 
mesons (left panel) detected in semi-central (20-60\%) Au+Au collisions at $\sqrt{s_{NN}} = 200$~GeV. 
The right panel show the same data scaled by quark number. These preliminary data
have been obtained by the PHENIX collaboration. 
}  
\label{phi_v2_ket}
\end{figure}

	The elliptic flow measurements for the $\phi$ meson are made difficult due to 
combinatoric background; for the case presented here, such a background results from $K^+K^{-}$ 
pairs which do not come from the decay of the $\phi$. Figures \ref{Phi_invmass}{\bf a} 
and {\bf b} show a typical example of invariant mass distributions  for K$^+K^{-}$ 
pairs obtained in Au+Au collisions at  $\sqrt{s_{NN}} = 200$~GeV. These preliminary 
data show results before and after combinatorial background subtraction. 
Here, it is important to stress that the combinatorial background also exhibits 
an azimuthal anisotropy ($v_2$) which needs to be separated from that for the 
$\phi$ meson. This can be achieved via a technique recently developed by 
Borghini et al. \cite{Borghini:2004ra}.  The technique which is based on  the 
study of $v_2$ as a function of inavariant mass, allows for robust $v_2$ extraction 
even for a rather small signal-to-background ratio. Needless to say, the distributions 
shown in Fig~\ref{Phi_invmass} allow a very reliable $v_2$ measurement for the $\phi$
meson.

	The left and right panels of Fig. \ref{phi_v2_ket} compares the unscaled and scaled 
results (respectively) for $v_2$ vs. $KE_T$ for $\pi, K, p$ and $\phi$ in 20-60\% 
central Au+Au collisions at  $\sqrt{s_{NN}} = 200$~GeV. These are recent preliminary
results obtained by the PHENIX collaboration. 
The left panel clearly shows that, despite its mass which 
is comparable to that for the proton, $v_2(KE_T)$ for the $\phi$ follows the flow 
pattern of the other light mesons ($\pi$ and $K$) whose cross sections are not 
OZI suppressed. The right panel indicate the expected universal scaling after $n_q$ scaling. 
We therefore conclude that, when elliptic flow develops, the constituents of 
the flowing medium can not be ground state hadrons interacting with their 
standard hadronic cross sections. Instead, they reflect a pre-hardonization 
state of the created high energy density matter that contain the prerequisite 
quantum numbers of the hadrons to be formed ie. the QGP. It is noteworthy 
that a similar investigation can be made via $v_2$ measurements 
for ``heavy" multi-strange baryons such as the $\Omega+\bar{\Omega}$; they too, are 
expected to have a relatively small hadronic scattering cross section. 
Indeed, we have observed that the recent high-statistics preliminary STAR results for 
$\Omega+\bar{\Omega}$ $v_2$ in Au+Au collisions at $\sqrt{s_{NN}} = 200$~GeV 
(presented at SQM 2006) \cite{Lamont:2006rc,Oldenburg:2006br} do follow the universal
$v_2/n_{q}$ vs $KE_T/n_{q}$ scaling (see Fig~ \ref{scalling_fig_ketn}).

\subsection{Validation of charm quark diffusion and flow  }
\begin{figure}[tb]
\includegraphics[width=1.\linewidth]{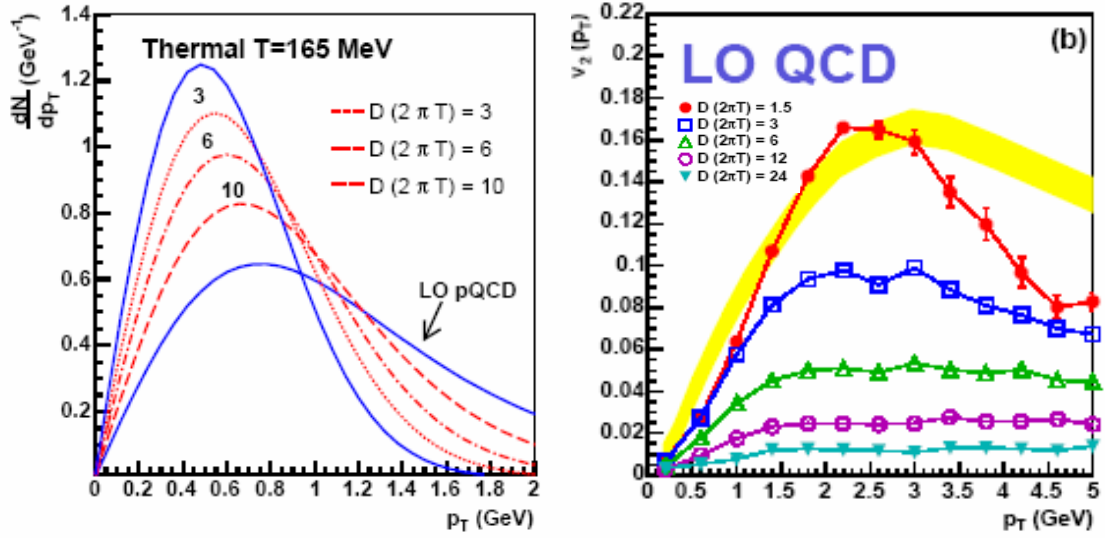}
%\vskip -1.2cm
\caption{ (Color on line) (a) Evolution of charm quark transverse momentum spectrum 
with different values of the diffusion coefficient $D_c$ as indicated. The thermal spectrum for 
$T = 165$ MeV and the initial transverse momentum spectrum given by leading order 
perturbation theory (LO pQCD) are indicated. (b) Evolution of charm quark $v_2(p_T)$ 
with $D_c$. These figures are taken from Ref. \cite{Moore:2004tg}.
}  
\label{charm_diff_Teaney}
\end{figure}

	As indicated earlier, perfect fluid hydrodynamics stipulates that all 
constituents of the QGP should experience the same velocity field. 
This requirement places severe constraints on the flow of heavy quarks 
and their attendant hadrons because their relaxation time $\tau_{R}$
is much larger than that for light quarks \cite{Moore:2004tg};
\[
	\tau_{R} \sim \frac{M}{T}\Gamma_s,
\]
where $M$ is the mass of the heavy quark, $T$ is the temperature and 
$\Gamma_s = \eta/sT$ is an estimate of the light quark relaxation time;
$\eta$ and $s$ are the viscosity and entropy density respectively. 
For the charm quark ($M \sim 1.4$ GeV) at a temperature of 165 MeV, 
this translates into an $\sim 8$ fold increase compared to $\Gamma_s$. 
Heavy quarks are also produced with a power-law or non-thermal transverse momentum
spectrum. Thus, one might expect the elliptic flow for heavy quarks to be much smaller 
than that for light quarks (for reasonable system lifetimes), unless in-medium 
interactions are strong enough to enforce many interactions with sufficient 
momentum transfer to thermalize them.

	The heavy quark diffusion coefficient $D_{hq}$, is related 
to $\tau_{R}$ \cite{Moore:2004tg};
\[
	\tau_{R} = \frac{M}{T}D_{hq}
\]	
Thus, $D_{hq}$ ultimately controls the extent to which the initial power-law spectrum approaches 
the thermal spectrum and the extent to which the heavy quark will follow the underlying 
flow of the medium. This is illustrated for the charm quark in 
Fig. \ref{charm_diff_Teaney} \cite{Moore:2004tg} where the evolution of the $p_T$ 
distribution and the differential $v_2(p_T)$, are shown for different values 
of the charm quark diffusion coefficient $D_c$. The figure clearly indicates that 
measurements of charm quark differential elliptic flow $v_2(p_T)$ can serve to 
constrain $D_c$. 
\begin{figure}[htb]
\includegraphics[width=1.\linewidth]{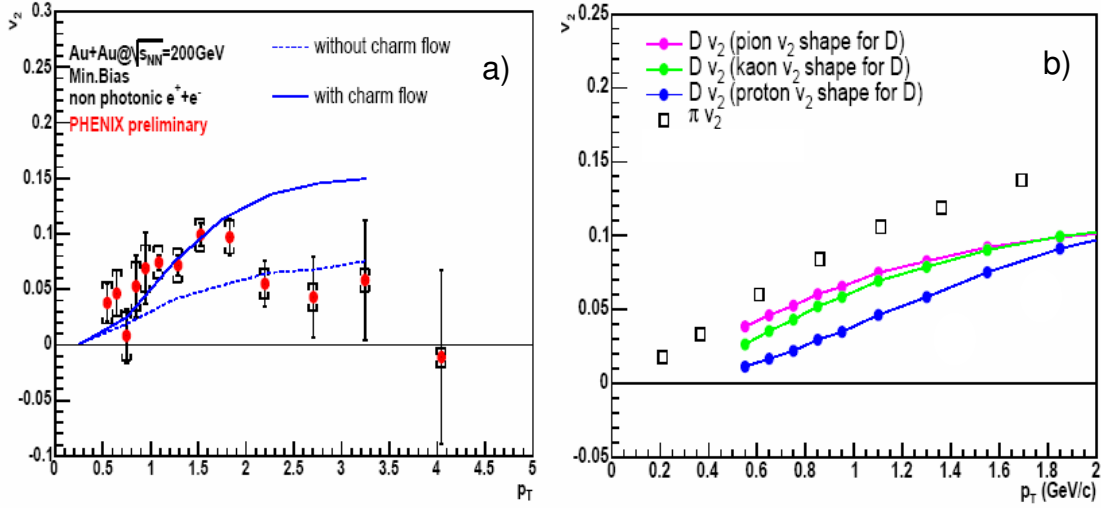}
%\vskip -1.2cm
\caption{ (Color on line) (a) $v_2$ vs. $p_T$ for non-photonic
electrons. The solid and dashed lines compare the results from a 
calculation which includes/excludes charm flow \cite{rapp_qm05}.
(b) $v_2$ vs. $p_T$ for D mesons derived from the non-photonic
electron measurements for $p_T < 2.0$ GeV/c. 
Results are shown for several assumptions for the shape of the $p_T$ 
dependence of the D meson $v_2$. The results for pions are shown for 
comparison.
}  
\label{charm_flow_phenix}
\end{figure}

	One method of accessing charm $v_2$ is to measure the the elliptic flow 
of ``non-photonic" electrons that originate primarily from the 
semi-leptonic decays of D and B mesons. Such a measurement has been made 
recently and found to yield significant anisotropy for D mesons \cite{Sakai:2005qn}.

	Figures \ref{charm_flow_phenix}a and b show differential $v_2(p_T)$ results 
obtained for non-photonic electrons and D mesons respectively. The D meson results 
reflect estimates for $p_T < 2.0$ GeV/c, obtained via detailed simulations constrained 
by the differential measurement $v_2(p_T)$ for non-photonic electrons \cite{Sakai:2005qn}.  
For these simulations, the $p_T$ 
spectrum was constrained by data and all non-photonic electrons were assumed 
to come from D meson decay  \cite{Sakai:2005qn}. Estimates were obtained for three separate 
{\bf shape} assumptions for $v_2(p_T)$ for the D mesons. The result from each 
is show in \ref{charm_flow_phenix}b. 

	These results indicate clear evidence that the D meson and consequently the 
charm quark, do flow with a large $v_2$. This suggest that the in-medium 
interactions are strong and perhaps frequent enough to thermalize the 
charm quark. If this is indeed the case, then $v_2(KE_T)$ for the D meson should 
show the same universal scaling presented earlier for other hadrons.
\begin{figure}[tb]
\includegraphics[width=1.\linewidth]{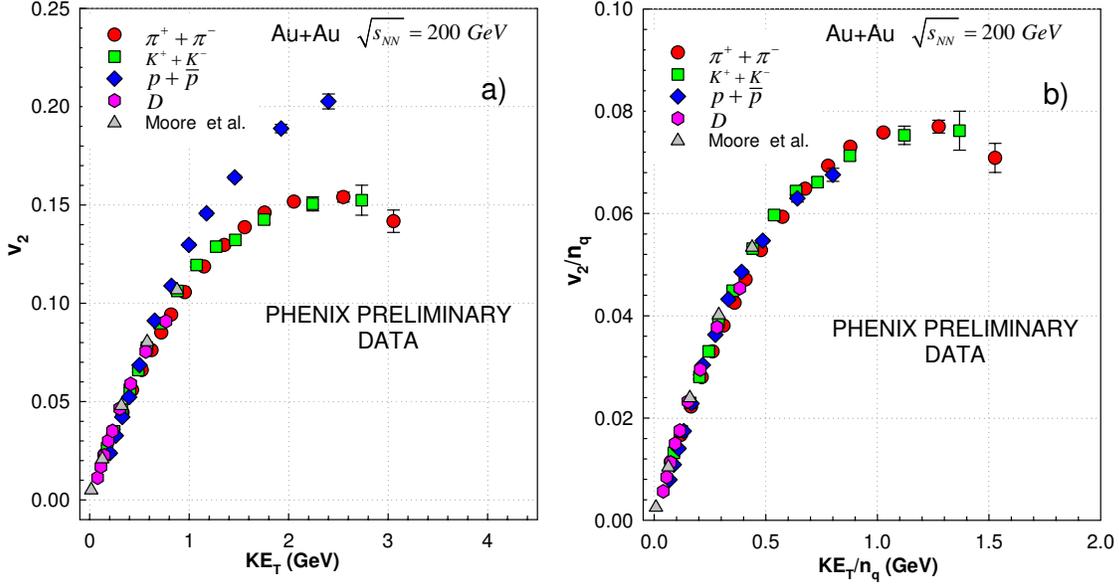}
%\vskip -1.2cm
\caption{ (Color on line) (a) $v_2$ vs. $KE_T$. (b) $v_2/n_q$ vs. $KE_T/n_q$. 
Results are shown for several particle species produced in minimum bias Au+Au 
collisions at $\sqrt{s_{NN}} = 200$ GeV. Interpolated results from a model 
calculation \cite{Moore:2004tg} for charm quarks are also indicated.
}  
\label{charm_flow_scaled_phenix}
\end{figure}

	Figures \ref{charm_flow_scaled_phenix}a and b  compare the unscaled and scaled 
results (respectively) for $v_2$ vs. $KE_T$ for $\pi, K, p$ and $D$ mesons measured 
in minimum bias  Au+Au collisions. The assumed shape for the transverse momentum 
dependence of the D meson $v_2$ is that for the proton cf. Fig. \ref{charm_flow_phenix}b.
We note here that this assumption is compatible with a large array of differential
flow measurements for different ``heavy" particle species 
(see for example Refs. \cite{Adams:2003am,Adams:2005zg,Lamont:2006rc,Oldenburg:2006br}).  
Both panels (a and b) of Fig. \ref{charm_flow_scaled_phenix} show robust scaling; they confirm 
that the D mesons (for $p_T < 2.0$ GeV/c) and their associated charm quarks flow with 
a fluid velocity field identical to that of the other constituents. 
This means that the medium responds as a thermalized fluid and the 
transport mean free path is small.

	The filled triangles in Fig. \ref{charm_flow_scaled_phenix} show results from an 
approximate interpolation of the model estimates for the charm quark \cite{Moore:2004tg}  
presented in Fig. \ref{charm_diff_Teaney}b. In Fig. \ref{charm_flow_scaled_phenix} these 
estimates are shown only  for $p_T \le 2.0$~GeV/c ie. the available range of the D meson 
data\footnote{The indicated scaling is observed to break for charm quarks with  
$p_T > 2.0$ GeV/c. Thus, high $p_T$ measurements are critical to the development 
of further constraints.}; they are close to the ones obtained from the LO-QCD charm quark 
spectrum for a calculation in which $ D_c \sim 3/{2\pi T}$ (see. Fig. \ref{charm_diff_Teaney}b). 
We note here that the calculations are for the charm quark and do not include 
the effects of hadronization/fragmentation leading to D meson production. Despite 
the many caveats of the model \cite{Moore:2004tg}, Fig. \ref{charm_flow_scaled_phenix} shows
that $v_2$ for the charm quark can be made to follow universal scaling for an appropriate 
choice of the coefficient $D_c$. It is noteworthy that an even smaller 
value $D_c \sim 1.5/{2\pi T}$ is found if the drag on the charm quark is made to 
depend on the velocity ie. $dp/dt \propto v$.

\section{Transport properties of the high energy density fluid }

	The observed scaling of elliptic flow provide important constraints for the 
transport properties of the high energy density QGP fluid. In what follows, we apply 
several of these constraints to make estimates for a number transport coefficients.

\subsection{Time scale for the development of elliptic flow }

	As discussed earlier, elliptic flow develops gradually in the 
evolving system on a time scale $\tau$ characterized by its transverse 
size $\bar R$ and the sound speed $c_s$:
\begin{equation}
	\tau \sim \frac{\bar R}{c_s}. 
\end{equation}	
Values for $\bar R$ have been reported as a function of centrality  
in Ref. \cite{Bhalerao:2005mm} for Au+Au collisions at $\sqrt{s_{NN}} = 60 - 200$ GeV. 
They indicate little change over the centrality 
range of interest here, so we use an averaged value $\left\langle \bar R \right\rangle\sim 1.9$. 
The value $\left\langle c_s\right\rangle =0.35 \pm 0.05$ determined via eccentricity 
scaling was presented earlier. These values 
give $\left\langle \tau \right\rangle \sim \left\langle \bar R \right\rangle/\left\langle c_s \right\rangle \sim 5.4$~fm, 
ie. the mean time required to ``translate" the initial spatial anisotropy of the high energy 
density fluid into a momentum anisotropy. 

\subsection{The viscosity to entropy ratio $\eta/s$  }
	
	Validation of the predictions from perfect fluid hydrodynamics for the 
scaling of the elliptic flow demands a low shear viscosity to entropy 
ratio ($\eta/s$) for the high energy density matter created. 
On the one hand, this is very well supported by the observation that experimental $v_2$ values 
are in fairly good agreement with simulated values based upon ideal 
hydrodynamics \cite{Shuryak:2004cy,Heinz:2001xi,Huovinen:2001cy,Hirano:2005wx}. On the  
hand the predicted suppression of flow, due to shear viscosity,  
by weak-coupling transport calculations is not observed. However, $\eta/s$ cannot be 
arbitrarily small because quantum mechanics limits the size of cross sections via 
unitarity. An absolute lower bound is $1/4\pi$, reached in the strong coupling limit 
of certain gauge theories. This bound has been recently conjectured to hold for all 
substances \cite{Kovtun:2004de}.

	The observed universal scaling of $v_2$ can be used to constrain a reliable 
estimate for $\eta/s$ \cite{Lacey:2006qb,Lacey:2006bc}:
\begin{equation}
	\frac{\eta}{s} \sim T \lambda_{f} c_s,
\end{equation}
where $T$ is the temperature, $\lambda_{f}$ is the mean free path and $c_s$ 
is the sound speed in the matter. The temperature $T=165\pm 3$~MeV was obtained via a 
hydrodynamically inspired fit to the scaled data shown in Fig.~\ref{scalling_fig_ketn}.
The functional form used for the fit is $I_1(w)/I_0(w)$, where $w=KE_T/2T$ 
and $I_1(w)$ and $I_0(w)$ are Bessel functions \cite{Csanad:2006sp}. 
For $c_s$, we use the estimate $c_s = 0.35 \pm 0.05$, given earlier and in 
Refs.~\cite{Issah:2006qn} and \cite{Adare:2006ti}. 
The mean free path estimate $\lambda_f = 0.3 \pm 0.03 fm$, was obtained
from an on-shell transport model simulation of the gluon evolution (in space 
and time) in Au+Au collisions at 200 GeV \cite{Xu:2004mz}. 
This parton cascade model includes pQCD $2 \leftrightarrow 2$ and 
$2 \leftrightarrow 3$ scatterings. These values for $\lambda_{f}$, $T$ and $c_s$ 
lead to the estimate $\eta/s \sim 0.09 \pm 0.015$ with an estimated systematic 
error of $ +0.1$ primarily due to uncertainties in $\lambda_f$. 

	This estimated value for $\eta/s$ is in good agreement with the 
experimentally based estimates of Teaney and Gavin \cite{Teaney:2003kp,Gavin:2006xd} 
and the theoretical estimates of Gyulassy and Shuryak \cite{Hirano:2005wx,Gelman:2006xw}. 
However, it is much lower than the value obtained from weak coupling QCD \cite{Arnold:2000dr} 
or hadronic computations \cite{Muronga:2003tb}. This low value for $\eta/s$ has been 
interpreted as evidence that the QGP is more strongly coupled than previously 
thought \cite{Shuryak:2004cy,Hirano:2005wx}.

\subsection{The bulk viscosity}

	Two component models (see for example Ref. \cite{Weinberg:1971m}) are often used to estimate 
the bulk viscosity $\varsigma$. Assuming such a model holds here, the bulk viscosity can be expressed in terms 
of the shear viscosity and the sound speed as:
\begin{equation}
    	\varsigma = \frac{5}{3}\left( 1 - 3c_s^2\right)^2.
\end{equation} 
Using the value $\left\langle c_s \right\rangle =0.35$, we find the 
relation $\left\langle \varsigma \right\rangle \sim 0.7\eta$. Thus, 
the implication is that the time averaged bulk viscosity of the high energy density matter is also
relatively small but non-zero. Here, it is important to stress  that for higher 
temperatures (eg. $T \sim 2T_c$), lattice QCD calculations give $c_s^2 \sim 1/3$ 
which leads to $\varsigma \sim 0.$ This implies that much of the bulk viscosity develops 
at the lower temperatures close to $T_c$.

\subsection{The sound attenuation length}

	With $T$ and $\eta/s$ in hand, the sound attenuation length $\Gamma_s$, can be evaluated as:
\begin{equation}
    \Gamma_s = \frac{4}{3}\frac{\eta}{sT}.
\end{equation} 
	This gives $ \Gamma_s \sim 0.16$~fm and $\eta/Ts \sim 0.12$~fm. These values 
suggests a rather short relaxation time for light quarks.

\subsection{The Reynolds $R_e$ and Knudsen $K_n$ numbers }

	The Reynolds $R_e$ and Knudsen $K_n$ numbers play an important role in hydrodynamic 
considerations;
\begin{equation}
	R_e \equiv \frac{\varepsilon \bar{R} c_s}{\eta}.
\end{equation}
Here $\varepsilon$ is the energy density. A large Reynolds number indicates that viscous 
forces are not important to the flow. That is, the smallest scales of fluid motion are 
undamped.

The Knudsen number is given as:
\begin{equation}
		K_n = \frac{\lambda}{\bar R}; 
\end{equation}
its reciprocal gives a measure of the number of collisions per particle. 
Thus, Local thermal equilibrium is expected if $K^{-1} >> 1$.  

	Since the shear viscosity $\eta$ can be expressed as:
\begin{equation}
	\eta = \varepsilon \lambda c_s,
\end{equation}
and flow is transonic (ie. Mach number $M_n \sim 1$),
\begin{equation}
	R_e \sim \frac{1}{K_n}.
\end{equation}
Using the previously obtained values for $\lambda$ and $\bar R$ we obtain 
$K^{-1} \sim 6$. Such a large number favors rapid local thermalization.

\subsection{Diffusion coefficient $D_c$ for the charm quark}

	The diffusion coefficient $D_c$, for the charm quark is related to 
its relaxation time $\tau_{R}$ \cite{Moore:2004tg};
\begin{equation}
	\tau_{R} = \frac{M}{T}D_c.
\end{equation}	
As discussed earlier, it controls the extent to which the initial power-law spectrum for 
the charm quark approaches the thermal distribution and the extent to which it follows 
the underlying flow of the medium. 

	The agreement between theory and experiment 
shown in Fig. \ref{charm_flow_scaled_phenix}, was achieved for 
\begin{equation}
	D_c \sim \frac{3}{2\pi T}. 
\end{equation}
This gives the values $D_c \sim 0.6 $~fm and $\tau_R \sim 5$~fm.
Thus, $D_c$ is approximately a factor of five times larger than 
the hydrodynamic scale $\eta/sT$ but $\tau_R$ is similar to the 
time for the development of elliptic flow. 
%It is noteworthy that if 
%one assumes a drag on the charm quark which 
%depends on the velocity ie. $dp/dt \propto v$, an even smaller value 
%$D_c \sim \frac{1.5}{2\pi T}$ is obtained.

\subsubsection{Does the value of $\eta/s$ signal the QCD critical end point?}
%

%
%Viscosity estimates :%%%
%
%
 \begin{figure}[tb]
 \begin{center}
 \includegraphics[width=0.5\linewidth]{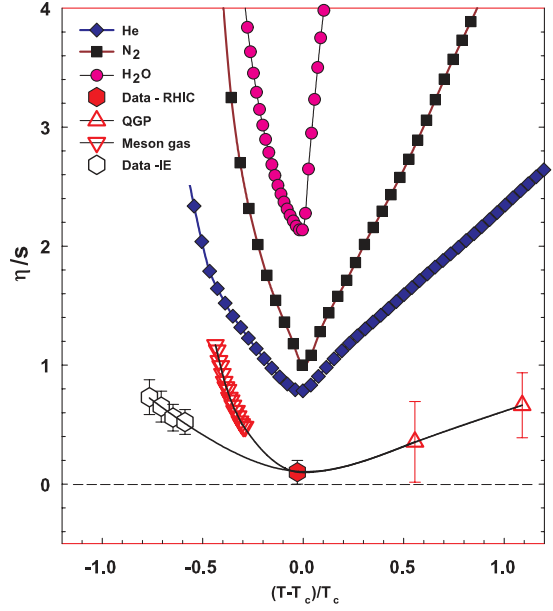}  
 \caption{\label{eta_over_s}
	(Color online) $\eta/s$ vs $(T-T_c)/T_c$ for several substances as indicated.
	The calculated values for the meson-gas have an associated error 
	of $\sim$ 50\%~\cite{Chen:2006ig}. 
	The lattice QCD value $T_c = 170$~MeV~\cite{Karsch:2000kv} 
	is assumed for nuclear matter. The lines are drawn to guide the eye.
}
\end{center}
 \end{figure}

Our earlier estimate of the viscosity to entropy ratio
is indicated by the filled hexagon is shown in Fig.~\ref{eta_over_s}. 
This figure shows  $\eta/s$ vs $(T-T_c)/T_c$ for molecular, atomic and nuclear 
matter. The data for He, N$_2$ and H$_2$O were obtained for their respective 
critical pressure. They are taken from Ref.~\cite{Csernai:2006zz} and the 
references therein. The calculated results shown for the meson-gas (for $T < T_c$) 
are obtained from chiral perturbation theory with free cross 
sections \cite{Chen:2006ig}. Those for the QGP (ie. for $T > T_c$)  
are from lattice QCD simulations \cite{Nakamura:2004sy}. 
The value $T_c \sim 170$ MeV is taken from lattice QCD 
calculations \cite{Karsch:2000kv}. 

	Figure~\ref{eta_over_s} illustrates the observation that for 
atomic and molecular substances, the ratio $\eta/s$ exhibits a minimum of 
comparable depth for isobars passing in the vicinity of the liquid-gas
critical point \cite{Kovtun:2004de,Csernai:2006zz}. When an isobar passes 
through the critical point (as shown in Fig.~\ref{eta_over_s}), the minimum 
forms a cusp at $T_c$; when it passes below the critical point, the minimum 
is found at a temperature below $T_c$ (liquid side) but is accompanied by a 
discontinuous change across the phase transition. For an isobar passing 
above the critical point, a less pronounced minimum is found at a value 
slightly above $T_c$. The value $\eta/s$ is smallest in the vicinity of $T_c$ 
because this corresponds to the most difficult condition for the transport of 
momentum \cite{Csernai:2006zz}.

	Given these observations, one expects a broad range of trajectories in 
the $(T,n_B)$ or $(T,\mu_B)$ plane for nuclear matter, to show $\eta/s$ minima 
with a possible cusp at the critical point. 
The exact location of this point is of course not known, and only coarse 
estimates of where it might lie are available. The open triangles in 
the figure show calculated values 
for $\eta/s$ along the $\mu_B=0$, $n_B=0$ trajectory. For $T < T_c$ the $\eta/s$ 
values for the meson-gas show an increase for decreasing values of $T$. 
For $T$ greater than $T_c$, the lattice results \cite{Nakamura:2004sy} 
indicate an increase of $\eta/s$ with $T$, albeit with large error bars.
These trends suggest a minimum for $\eta/s$ in the 
vicinity of $T_c$. This minimum is rather close to the absolute lower bound 
of $\eta/s = 1/4\pi$. We therefore speculate that 
it is compatible with the minimum expected if the hot and dense QCD matter 
produced in RHIC collisions follow decay trajectories which are close to 
the QCD critical end point (CEP). Such trajectories could be readily 
followed if the CEP acts as an attractor for thermodynamic trajectories of 
the decaying matter \cite{Nonaka:2004pg,Kampfer:2005nt}.

\section{Summary} 

In summary, elliptic flow measurements at RHIC provide compelling evidence for the 
production and rapid thermalization 
(less than 1 fm/c after impact) of a new state of matter having an energy density well 
in excess of the critical value required for de-confinement i.e the QGP. Validation of the 
predictions from perfect fluid hydrodynamics for the scaling of the elliptic 
flow, in conjunction with the universal scaling of baryons and mesons (via quark 
number), serve as constraints for the extraction of several transport coefficients 
for this matter. These coefficients give important insights on the transport 
properties and the decay dynamics of the QGP and suggest a 
%indicate a new state of matter a consistent with that of a strongly coupled 
plasma having essentially perfect liquid-like properties. 
%
\input{references}

%\bibliography{CorFluc06.bib}
%
%
\end{document}

%% file: references.tex
%

%% file: CorFluc06_Lacey3.bbl
\begin{thebibliography}{9}
%

%\cite{Adcox:2004mh}
\bibitem{Adcox:2004mh}
  K.~Adcox {\it et al.}  [PHENIX Collaboration],
  % ``Formation of dense partonic matter in relativistic nucleus nucleus
  %collisions at RHIC: Experimental evaluation by the PHENIX  collaboration,''
  Nucl.\ Phys.\ A {\bf 757} (2005) 184
  %%CITATION = NUCL-EX 0410003;%%


%\cite{Karsch:2001vs}
\bibitem{Karsch:2001vs}
  F.~Karsch,
  %``Lattice results on QCD thermodynamics,''
  Nucl.\ Phys.\ A {\bf 698} (2002) 199
  %%CITATION = HEP-PH 0103314;%%


\bibitem{Fodor:2001pe}
  Z.~Fodor and S.~D.~Katz,
  %``Lattice determination of the critical point of QCD at finite T and mu,''
  JHEP {\bf 0203} (2002) 014
  %%CITATION = HEP-LAT 0106002;%%

%\cite{Adams:2005dq}
\bibitem{Adams:2005dq}
  J.~Adams {\it et al.}  [STAR Collaboration],
  % ``Experimental and theoretical challenges in the search for the quark  gluon
  % plasma: The STAR collaboration's critical assessment of the  evidence from
  %RHIC collisions,''
  Nucl.\ Phys.\ A {\bf 757} (2005) 102
  %%CITATION = NUCL-EX 0501009;%%


%\cite{Back:2004je}
\bibitem{Back:2004je}
  B.~B.~Back {\it et al.}, [PHOBOS Collaboration],
  %``The PHOBOS perspective on discoveries at RHIC,''
  Nucl.\ Phys.\ A {\bf 757} (2005) 28
  %%CITATION = NUCL-EX 0410022;%%


%\cite{Arsene:2004fa}
\bibitem{Arsene:2004fa}
  I.~Arsene {\it et al.}  [BRAHMS Collaboration],
  % ``Quark gluon plasma and color glass condensate at RHIC? The perspective
  %from the BRAHMS experiment,''
  Nucl.\ Phys.\ A {\bf 757} (2005) 1
  %%CITATION = NUCL-EX 0410020;%%

%\cite{Lacey:2005qq}
\bibitem{Lacey:2005qq}
  R.~A.~Lacey,
  % ``The role of elliptic flow correlations in the discovery of the sQGP at
  %RHIC,''
  Nucl.\ Phys.\ A {\bf 774} (2006) 199
  %%CITATION = NUCL-EX 0510029;%%


%\cite{Roland:2005ei}
\bibitem{Roland:2005ei}
  G.~Roland {\it et al.}  [PHOBOS Collaboration],
  %``New results from the PHOBOS experiment,''
  Nucl.\ Phys.\ A {\bf 774} (2006) 113
  %%CITATION = NUCL-EX 0510042;%%


%\cite{Braun-Munzinger:2001ip}
\bibitem{Braun-Munzinger:2001ip}
  P.~Braun-Munzinger, D.~Magestro, K.~Redlich and J.~Stachel,
  %``Hadron production in Au Au collisions at RHIC,''
  Phys.\ Lett.\ B {\bf 518} (2001) 41
  %%CITATION = HEP-PH 0105229;%%

%\cite{Adler:2003kt}
\bibitem{Adler:2003kt}
  S.~S.~Adler {\it et al.}  [PHENIX Collaboration],
  % ``Elliptic flow of identified hadrons in Au + Au collisions at  s(NN)**(1/2)
  %= 200-GeV,''
  Phys.\ Rev.\ Lett.\  {\bf 91} (2003) 182301
  %%CITATION = NUCL-EX 0305013;%%


%\cite{Adams:2003am}
\bibitem{Adams:2003am}
  J.~Adams {\it et al.}  [STAR Collaboration],
  % ``Particle dependence of azimuthal anisotropy and nuclear modification of
  % particle production at moderate p(T) in Au + Au collisions at  s(NN)**(1/2) =
  %200-GeV,''
  Phys.\ Rev.\ Lett.\  {\bf 92} (2004) 052302
  %%CITATION = NUCL-EX 0306007;%%


%\cite{Adams:2004bi}
\bibitem{Adams:2004bi}
  J.~Adams {\it et al.}  [STAR Collaboration],
  %``Azimuthal anisotropy in Au + Au collisions at s(NN)**(1/2) = 200-GeV,''
  Phys.\ Rev.\ C {\bf 72} (2005) 014904
  %%CITATION = NUCL-EX 0409033;%%


%\cite{Adams:2005zg}
\bibitem{Adams:2005zg}
  J.~Adams {\it et al.}  [STAR Collaboration],
  % ``Multi-strange baryon elliptic flow in Au + Au collisions at  s(NN)**(1/2) =
  %200-GeV,''
  Phys.\ Rev.\ Lett.\  {\bf 95} (2005) 122301
  %%CITATION = NUCL-EX 0504022;%%


%\cite{Sakai:2005qn}
\bibitem{Sakai:2005qn}
  S.~Sakai, [PHENIX Collaboration],
  % ``The azimuthal anisotropy of electrons from heavy flavor decays in
  %s(NN)**(1/2) = 200-GeV Au - Au collisions by PHENIX,''
  nucl-ex/0510027.
  %%CITATION = NUCL-EX 0510027;%%

%\cite{Voloshin:1994mz}
\bibitem{Voloshin:1994mz}
  S.~Voloshin and Y.~Zhang,
  % ``Flow study in relativistic nuclear collisions by Fourier expansion of
  %Azimuthal particle distributions,''
  Z.\ Phys.\ C {\bf 70} (1996) 665
  %%CITATION = HEP-PH 9407282;%%



%\cite{Danielewicz:1985hn}
\bibitem{Danielewicz:1985hn}
P.~Danielewicz and G.~Odyniec,
%``Transverse Momentum Analysis Of Collective Motion In Relativistic Nuclear Collisions,''
Phys.\ Lett.\ B {\bf 157}, 146 (1985).
%%CITATION = PHLTA,B157,146;%%

%\cite{Poskanzer:1998yz}
\bibitem{Poskanzer:1998yz}
A.~M.~Poskanzer and S.~A.~Voloshin,
%``Methods for analyzing anisotropic flow in relativistic nuclear  collisions,''
Phys.\ Rev.\ C {\bf 58}, 1671 (1998)
[nucl-ex/9805001].
%%CITATION = NUCL-EX 9805001;%%



%\cite{Ollitrault:1997di}
\bibitem{Ollitrault:1997di}
J.-Y.~Ollitrault,
%``On the measurement of azimuthal anisotropies in nucleus--nucleus collisions,''
nucl-ex/9711003.
%%CITATION = NUCL-EX 9711003;%%



%\cite{Ollitrault:1992bk}
\bibitem{Ollitrault:1992bk}
  J.~Y.~Ollitrault,
  %``Anisotropy As A Signature Of Transverse Collective Flow,''
  Phys.\ Rev.\ D {\bf 46} (1992) 229.
  %%CITATION = PHRVA,D46,229;%%


%\cite{Heiselberg:1998es}
\bibitem{Heiselberg:1998es}
  H.~Heiselberg and A.~M.~Levy,
  %``Elliptic flow and HBT in non-central nuclear collisions,''
  Phys.\ Rev.\ C {\bf 59} (1999) 2716
  %%CITATION = NUCL-TH 9812034;%%





%\cite{Huovinen:2001cy}
\bibitem{Huovinen:2001cy}
  P.~Huovinen and others,
  %``Radial and elliptic flow at RHIC: Further predictions,''
  Phys.\ Lett.\ B {\bf 503} (2001) 58
  [arXiv:hep-ph/0101136].
  %%CITATION = HEP-PH 0101136;%%
  
\bibitem{teaney2001}
  D.~Teaney, J.~Lauret, and E.V.~Shuryak, nucl-th/ 0011058.

\bibitem{kolb2001} P.\ F.\ Kolb {\em et al.}, hep-ph/0103234

\bibitem{hirano_qm05} T.\ Hirano {\em et al.}, these proceedings.

%\cite{Bhalerao:2005mm}
\bibitem{Bhalerao:2005mm}
  R.~S.~Bhalerao, J.~P.~Blaizot, N.~Borghini and J.~Y.~Ollitrault,
  %``Elliptic flow and incomplete equilibration at RHIC,''
  Phys.\ Lett.\ B {\bf 627} (2005) 49
  %%CITATION = NUCL-TH 0508009;%%



%\cite{Karsch:2006sm}
\bibitem{Karsch:2006sm}
  F.~Karsch,
  %``Lattice QCD at high temperature and the QGP,''
  arXiv:hep-lat/0601013.
  %%CITATION = HEP-LAT 0601013;%%
%\cite{Adler:2005rg}
\bibitem{Adler:2005rg}
  S.~S.~Adler {\it et al.}  [PHENIX Collaboration],
  %``Measurement of identified pi0 and inclusive photon v(2) and implication  to
  %the direct photon production in s(NN)**(1/2) = 200-GeV Au + Au  collisions,''
  Phys.\ Rev.\ Lett.\  {\bf 96} (2006) 032302
  %%CITATION = NUCL-EX 0508019;%%


%\cite{Issah:2006qn}
\bibitem{Issah:2006qn}
  M.~Issah and A.~Taranenko  [PHENIX Collaboration],
  %``Scaling characteristics of azimuthal anisotropy at RHIC,''
  arXiv:nucl-ex/0604011.
  %%CITATION = NUCL-EX 0604011;%%


%\cite{Adare:2006ti}
\bibitem{Adare:2006ti}
  A.~Adare  [PHENIX Collaboration],
  % ``Scaling properties of azimuthal anisotropy in Au + Au and Cu + Cu
  %collisions at s(NN)**(1/2) = 200-GeV,''
  arXiv:nucl-ex/0608033.
  %%CITATION = NUCL-EX 0608033;%%

%\cite{Oldenburg:2006br}
\bibitem{Oldenburg:2006br}
  M.~Oldenburg  [STAR Collaboration],
  % ``Centrality dependence of azimuthal anisotropy of strange hadrons in 200-GeV
  %Au + Au collisions,''
  arXiv:nucl-ex/0607021.
  %%CITATION = NUCL-EX 0607021;%%

%\cite{Lamont:2006rc}
\bibitem{Lamont:2006rc}
  M.~A.~C.~Lamont  [STAR Collaboration],
  %``Recent results on strangeness production at RHIC,''
  arXiv:nucl-ex/0608017.
  %%CITATION = NUCL-EX 0608017;%%


%\cite{Csanad:2005gv}
\bibitem{Csanad:2005gv}
  M.~Csanad {\it et al.},
  % ``Universal scaling of the elliptic flow and the perfect hydro picture at
  %RHIC,''
  arXiv:nucl-th/0512078.
  %%CITATION = NUCL-TH 0512078;%%


%\cite{Borghini:2005kd}
\bibitem{Borghini:2005kd}
  N.~Borghini and J.~Y.~Ollitrault,
  %``Momentum spectra, anisotropic flow, and ideal fluids,''
  arXiv:nucl-th/0506045.
  %%CITATION = NUCL-TH 0506045;%%




%\cite{Csanad:2006sp}
\bibitem{Csanad:2006sp}
  M.~Csanad, T.~Csorgo, R.~A.~Lacey and B.~Lorstad,
  %``Universal scaling of the elliptic flow at RHIC,''
  arXiv:nucl-th/0605044.
  %%CITATION = NUCL-TH 0605044;%%


%\cite{Voloshin:2002wa}
\bibitem{Voloshin:2002wa}
  S.~A.~Voloshin,
  %``Anisotropic flow,''
  Nucl.\ Phys.\ A {\bf 715} (2003) 379
  %%CITATION = NUCL-EX 0210014;%%



\bibitem{Fries:2003kq}
  R.~J.~Fries, B.~Muller, C.~Nonaka and S.~A.~Bass,
  %``Hadron production in heavy ion collisions: Fragmentation and  recombination
  %from a dense parton phase,''
  Phys.\ Rev.\ C {\bf 68} (2003) 044902
  %%CITATION = NUCL-TH 0306027;%%


%\cite{Molnar:2003ff}
\bibitem{Molnar:2003ff}
  D.~Molnar and S.~A.~Voloshin,
  %``Elliptic flow at large transverse momenta from quark coalescence,''
  Phys.\ Rev.\ Lett.\  {\bf 91} (2003) 092301
  %%CITATION = NUCL-TH 0302014;%%

%\cite{Greco:2003mm}
\bibitem{Greco:2003mm}
  V.~Greco, C.~M.~Ko and P.~Levai,
  %``Parton coalescence at RHIC,''
  Phys.\ Rev.\ C {\bf 68} (2003) 034904
  %%CITATION = NUCL-TH 0305024;%%


%\cite{Adcox:2002ms}
\bibitem{Adcox:2002ms}
  K.~Adcox {\it et al.}  [PHENIX Collaboration],
  % ``Flow measurements via two-particle azimuthal correlations in Au + Au
  %collisions at s(NN)**(1/2) = 130-GeV,''
  Phys.\ Rev.\ Lett.\  {\bf 89} (2002) 212301
  %%CITATION = NUCL-EX 0204005;%%



%\cite{Shuryak:2004cy}
\bibitem{Shuryak:2004cy}
  E.~V.~Shuryak,
  % ``What RHIC experiments and theory tell us about properties of  quark-gluon
  %plasma?,''
  Nucl.\ Phys.\ A {\bf 750} (2005) 64
  %%CITATION = HEP-PH 0405066;%%



%\cite{Heinz:2001xi}
\bibitem{Heinz:2001xi}
  U.~W.~Heinz and P.~F.~Kolb,
  %``Early thermalization at RHIC,''
  Nucl.\ Phys.\ A {\bf 702} (2002) 269
  %%CITATION = HEP-PH 0111075;%%



%\cite{Adler:2004cj}
\bibitem{Adler:2004cj}
  S.~S.~Adler {\it et al.}  [PHENIX Collaboration],
  % ``Saturation of azimuthal anisotropy in Au + Au collisions at  s(NN)**(1/2) =
  %62-GeV - 200-GeV,''
  Phys.\ Rev.\ Lett.\  {\bf 94} (2005) 232302
  %%CITATION = NUCL-EX 0411040;%%



%\cite{Okubo:1963fa}
\bibitem{Okubo:1963fa}
  S.~Okubo,
  %``Phi meson and unitary symmetry model,''
  Phys.\ Lett.\  {\bf 5} (1963) 165.
  %%CITATION = PHLTA,5,165;%%


%\cite{Shor:1984ui}
\bibitem{Shor:1984ui}
  A.~Shor,
  %``Phi Meson Production As A Probe Of The Quark Gluon Plasma,''
  Phys.\ Rev.\ Lett.\  {\bf 54} (1985) 1122.
  %%CITATION = PRLTA,54,1122;%%


%\cite{Ko:1993id}
\bibitem{Ko:1993id}
  C.~M.~Ko and D.~Seibert,
  % ``What can we learn from a second phi meson peak in ultrarelativistic nuclear
  %collisions?,''
  Phys.\ Rev.\ C {\bf 49} (1994) 2198
  %%CITATION = NUCL-TH 9312010;%%



%\cite{Haglin:1994xu}
\bibitem{Haglin:1994xu}
  K.~Haglin,
  %``Collision rates for rho, omega and phi mesons at nonzero temperature,''
  Nucl.\ Phys.\ A {\bf 584} (1995) 719
  %%CITATION = NUCL-TH 9410028;%%


%\cite{Xu:2005jt}
\bibitem{Xu:2005jt}
  N.~Xu,
 %  ``Collective expansion in high-energy nuclear collisions: The search for the
  %partonic EOS at RHIC,''
  Nucl.\ Phys.\ A {\bf 751} (2005) 109.
  %%CITATION = NUPHA,A751,109;%%


%\cite{Borghini:2004ra}
\bibitem{Borghini:2004ra}
  N.~Borghini and J.~Y.~Ollitrault,
  %``Azimuthally sensitive correlations in nucleus nucleus collisions,''
  Phys.\ Rev.\ C {\bf 70} (2004) 064905
  %%CITATION = NUCL-TH 0407041;%%

%\cite{Moore:2004tg}
\bibitem{Moore:2004tg}
  G.~D.~Moore and D.~Teaney,
  %``How much do heavy quarks thermalize in a heavy ion collision?,''
  Phys.\ Rev.\ C {\bf 71} (2005) 064904
  %%CITATION = HEP-PH 0412346;%%

% 
\bibitem{rapp_qm05} 
   R. Rapp et al., hep-ph/0510050 (2005)

%\cite{Weinberg:1971m}
\bibitem{Weinberg:1971m}
  S. Weinberg,
  %
   Astrophys. J. , 168 (1971) 175
   
   
%\cite{Hirano:2005wx}
\bibitem{Hirano:2005wx}
  T.~Hirano and M.~Gyulassy,
  % ``Perfect fluidity of the quark gluon plasma core as seen through its
  %dissipative hadronic corona,''
  Nucl.\ Phys.\ A {\bf 769} (2006) 71
  %%CITATION = NUCL-TH 0506049;%%



%\cite{Kovtun:2004de}
\bibitem{Kovtun:2004de}
  P.~Kovtun, D.~T.~Son and A.~O.~Starinets,
  % ``Viscosity in strongly interacting quantum field theories from black hole
  %physics,''
  Phys.\ Rev.\ Lett.\  {\bf 94} (2005) 111601
  %%CITATION = HEP-TH 0405231;%%


%\cite{Lacey:2006qb}
\bibitem{Lacey:2006qb}
  R.~A.~Lacey,
  %``Is there a sonic boom in the little bang at RHIC?,''
  arXiv:nucl-ex/0608046.
  %%CITATION = NUCL-EX 0608046;%%


%\cite{Lacey:2006bc}
\bibitem{Lacey:2006bc}
  R.~A.~Lacey {\it et al.},
  %``Has the QCD critical point been signaled by observations at RHIC?,''
  arXiv:nucl-ex/0609025.
  %%CITATION = NUCL-EX 0609025;%%

%\cite{Xu:2004mz}
\bibitem{Xu:2004mz}
  Z.~Xu and C.~Greiner,
  % ``Thermalization of gluons in ultrarelativistic heavy ion collisions by
  %including three-body interactions in a parton cascade,''
  Phys.\ Rev.\ C {\bf 71} (2005) 064901
  %%CITATION = HEP-PH 0406278;%%


%\cite{Teaney:2003kp}
\bibitem{Teaney:2003kp}
  D.~Teaney,
  % ``Effect of shear viscosity on spectra, elliptic flow, and Hanbury
  %Brown-Twiss radii,''
  Phys.\ Rev.\ C {\bf 68} (2003) 034913
  %%CITATION = NUCL-TH 0301099;%%

%\cite{Gavin:2006xd}
\bibitem{Gavin:2006xd}
  S.~Gavin and M.~Abdel-Aziz,
  % ``Measuring Shear Viscosity Using Transverse Momentum Correlations in
  %Relativistic Nuclear Collisions,''
  arXiv:nucl-th/0606061.
  %%CITATION = NUCL-TH 0606061;%%


%\cite{Gelman:2006xw}
\bibitem{Gelman:2006xw}
  B.~A.~Gelman, E.~V.~Shuryak and I.~Zahed,
   %``Classical Strongly Coupled QGP I: The Model and Molecular Dynamics
  %Simulations,''
  arXiv:nucl-th/0601029.
  %%CITATION = NUCL-TH 0601029;%%

%\cite{Muronga:2003tb}
\bibitem{Muronga:2003tb}
  A.~Muronga,
  %``Shear Viscosity Coefficient from Microscopic Models,''
  Phys.\ Rev.\ C {\bf 69} (2004) 044901
  %%CITATION = NUCL-TH 0309056;%%

%\cite{Arnold:2000dr}
\bibitem{Arnold:2000dr}
  P.~Arnold, G.~D.~Moore and L.~G.~Yaffe,
  % ``Transport coefficients in high temperature gauge theories. I:  Leading-log
  %results,''
  JHEP {\bf 0011} (2000) 001
  %%CITATION = HEP-PH 0010177;%%

%\cite{Chen:2006ig}
\bibitem{Chen:2006ig}
  J.~W.~Chen and E.~Nakano,
  % ``Shear viscosity to entropy density ratio of QCD below the deconfinement
  %temperature,''
  arXiv:hep-ph/0604138.
  %%CITATION = HEP-PH 0604138;%%



%\cite{Karsch:2000kv}
\bibitem{Karsch:2000kv}
  F.~Karsch, E.~Laermann and A.~Peikert,
  %``Quark mass and flavor dependence of the QCD phase transition,''
  Nucl.\ Phys.\ B {\bf 605} (2001) 579
  [arXiv:hep-lat/0012023].
  %%CITATION = HEP-LAT 0012023;%%


%\cite{Nonaka:2004pg}
\bibitem{Nonaka:2004pg}
  C.~Nonaka and M.~Asakawa,
  %``Hydrodynamical evolution near the QCD critical end point,''
  Phys.\ Rev.\ C {\bf 71} (2005) 044904
  %%CITATION = NUCL-TH 0410078;%%

%\cite{Kampfer:2005nt}
\bibitem{Kampfer:2005nt}
  B.~Kampfer, M.~Bluhm, R.~Schulze, D.~Seipt and U.~Heinz,
  %``QCD matter within a quasi-particle model and the critical end point,''
  Nucl.\ Phys.\ A {\bf 774} (2006) 757
  %%CITATION = HEP-PH 0509146;%%

%\cite{Csernai:2006zz}
\bibitem{Csernai:2006zz}
  L.~P.~Csernai, J.~I.~Kapusta and L.~D.~McLerran,
  % ``On the strongly-interacting low-viscosity matter created in relativistic
  %nuclear collisions,''
  arXiv:nucl-th/0604032.
  %%CITATION = NUCL-TH 0604032;%%

%\cite{Nakamura:2004sy}
\bibitem{Nakamura:2004sy}
  A.~Nakamura and S.~Sakai,
  %``Transport coefficients of gluon plasma,''
  Phys.\ Rev.\ Lett.\  {\bf 94} (2005) 072305
  %%CITATION = HEP-LAT 0406009;%%


\end{thebibliography}
